\documentclass[11pt]{revtex4}    
\usepackage{amssymb,epsf}
\usepackage{latexsym}
\begin{document}

\title{Magnetic Branes Supported by Nonlinear Electromagnetic Field}
\author{S. H. Hendi\footnote{hendi@mail.yu.ac.ir}}

\address{Physics Department,
College of Sciences, Yasouj University, Yasouj
75914, Iran\\
National Elite Foundation, P.O. Box 19615-334, Tehran, Iran}

\begin{abstract}
Considering the nonlinear electromagnetic field coupled to Einstein gravity
in the presence of cosmological constant, we obtain a new class of $d$%
-dimensional magnetic brane solutions. This class of solutions yields a
spacetime with a longitudinal nonlinear magnetic field generated by a static
source. These solutions have no curvature singularity and no horizons but
have a conic geometry with a deficit angle $\delta \phi$. We investigate the
effects of nonlinearity on the metric function and deficit angle and also
find that for the special range of the nonlinear parameter, the solutions
are not asymptotic AdS. We generalize this class of solutions to the case of
spinning magnetic solutions, and find that when one or more rotation
parameters are nonzero, the brane has a net electric charge which is
proportional to the magnitude of the rotation parameters. Then, we use the
counterterm method and compute the conserved quantities of these spacetimes.
Finally, we obtain a constrain on the nonlinear parameter, such that the
nonlinear electromagnetic field is conformally invariant.
\end{abstract}

\maketitle

\section{Introduction}

Research on the topological defects have expanded tremendously during the
last decade and still remains one of the most active fields in modern
cosmology. Topological defects are inevitably formed during phase
transitions in the early universe, and their subsequent evolution and
observational signatures must therefore be understood. Also, apart from
their possible astrophysical roles, topological defects are fascinating
objects in their own right. Their properties, which are very different from
those of more familiar system, can give rise to a rich variety of unusual
mathematical and physical phenomena.

The string model of structure formation may help to resolve one of
cosmological mystery, the origin of cosmic magnetic fields \cite{VachVil91}.
There is strong evidence from all numerical simulations for the scaling
behavior of the long string network during the radiation-dominated era.
Despite the original lack of consensus on the actual long string density,
all simulations demonstrated that high or low initial string densities were
driven towards a stable fixed point \cite{VilenkinBook}.

Metrics describing the causal birth of a cosmic string have been studied by
Mageuijo \cite{Mage92}. In a specific toy model for straight string
formation, the metric takes account of a compensating radiation under
density and also an outgoing gravitational shock wave where the deficit
angle switches off.

Solutions of Einstein's equations with conical singularities describing
straight strings can easily be constructed \cite{AryForVil86}. One needs
only a spacetime with a symmetry axis. If one then cuts out a wedge then a
space with a string lying along the axis is obtained. A non axisymmetric
solution of the combined Einstein and Maxwell equations with a string has
been found by Linet \cite{Linet87}. This solution describes a maximally
charged black hole with an infinite string.

Today, nonlinear electromagnetic fields in higher dimensions are subjects of
interest in recent years. For example, there has been a renewed interest in
Born-Infeld gravity ever since new solutions have been found in the low
energy limit of string theory. Static and rotating solutions of Born-Infeld
gravity have been considered in Refs. \cite{Demi86,DehSed,DAH08DH07}.

In this paper we want to introduce magnetic solutions of the Einstein
gravity in the presence of a nonlinear electromagnetic field and find the
effects of the nonlinearity on the properties of the solutions. The basic
motivation of considering this spacetime is that near the origin, $4$%
-dimensional section of this metric may be interpret as a topological string
(in analogy with cosmic string). Also, the presented solutions is a
generalization of usual linear electromagnetic field and the nonlinearity
effects on the electromagnetic field, solutions, charge, deficit angle and
so on. In the other word the presented solutions are completely differ from
the usual linear electromagnetic field. As we see, our solutions do not have
the same asymptotic behavior with respect to linear case. It is interesting
to note that, in general case, the expression of the electric field depends
on the dimension, $d$, and the power of nonlinearity, $s$, simultaneously
and its value coincides with the higher dimensional Reissner-Nordstr\"{o}m
solutions for linear limit. These solutions are very flexible and there is
not any restriction on the nonlinear parameter $s$ and one can fix it for
consistency to other theoretical and experimental evidences. Solutions with
longitudinal and angular magnetic field were considered in Refs. \cite%
{Dehhorizonless1,Dehhorizonless2,DB07,DHWorm}. Similar static solutions in
the context of cosmic string theory were found in Ref. \cite{VilBanBanSen96}%
. All of these solutions \cite%
{Dehhorizonless1,Dehhorizonless2,DB07,VilBanBanSen96,LeviCivita19Marder58}
are horizonless and have a conical geometry; they are everywhere flat except
at the location of the line source. The extension to include the
electromagnetic field has also been done \cite{Mukh38Witt85,DiasLemos02}.
The generalization of these solutions in Einstein gravity in the presence of
a dilaton and Born-Infeld electromagnetic fields has been done in Ref. \cite%
{HendiJMP08DSH08}.

Another example of the nonlinear electromagnetic field is conformally
invariant Maxwell field. In many papers, straightforward generalization of
the Maxwell field to higher dimensions one essential property of the
electromagnetic field is lost, namely, conformal invariance. Indeed, in the
Reissner-Nordstr\"{o}m solution, the source is given by the Maxwell action
which enjoys the conformal invariance in four dimensions. Massless spin-$1/2$
fields have vanishing classical stress tensor trace in any dimension, while
scalars can be \textquotedblleft improved\textquotedblright\ to achieve $%
T_{\alpha }^{\alpha }=0$, thereby guaranteeing invariance under the special
conformal (or full Weyl) group, in accord with their scale-independence \cite%
{DesSch}. Maxwell theory can be studied in a gauge which is invariant under
conformal rescalings of the metric, and at first, have been proposed by
Eastwood and Singer \cite{EastSing}. Also, Poplawski \cite{Popl08} have been
showed the equivalence between the Ferraris--Kijowski and Maxwell Lagrangian
results from the invariance of the latter under conformal transformations of
the metric tensor. Quantized Maxwell theory in a conformally invariant gauge
have been investigated by Esposito \cite{Espo97}. In recent years, gravity
in the presence of nonlinear and conformally invariant Maxwell source have
been studied in many papers \cite{HessMart07MHMar,HendiRast}.

The outline of our paper is as follows. We give a brief review of the field
equations of Einstein gravity in the presence of cosmological constant and
nonlinear electromagnetic field in Sec. \ref{Fiel}. In Sec. \ref{Long} we
present static horizonless solutions which produce longitudinal magnetic
field, compare it with the solutions of the standard electromagnetic field
and then investigate the properties of the solutions and the effects of
nonlinearity of the electromagnetic field on the deficit angle of the
spacetime. Section \ref{Rot} will be devoted to the generalization of these
solutions to the case of rotating solutions and use of the counterterm
method to compute the conserved quantities of them. We finish our paper with
some concluding remarks.

\section{Field equations\label{Fiel}}

We consider the $d$-dimensional action in which gravity is coupled to
nonlinear electrodynamics field in the presence of negative cosmological
constant with an action
\begin{equation}
I_{G}\left( g_{\mu \nu },A_{\mu }\right) =-\frac{1}{16\pi }%
\int\limits_{\partial M}d^{d}x\sqrt{-g}\left[ R-2\Lambda -\alpha F^{s}\right]
,  \label{IG}
\end{equation}
where $R$ is the scalar curvature, $\Lambda =-(d-1)(d-2)/2l^{2}$ is the
negative cosmological constant for asymptotically AdS solutions, $F$ is the
Maxwell invariant which is equal to$\ F_{\mu \nu }F^{\mu \nu }$(where $%
F_{\mu \nu }=\partial _{\mu }A_{\nu }-\partial _{\nu }A_{\mu }$ is the
electromagnetic tensor field and $A_{\mu }$ is the vector potential), $%
\alpha $ and $s$\ is a coupling and positive arbitrary constant
respectively. Varying the action (\ref{IG}) with respect to the metric
tensor $g_{\mu \nu }$ and the electromagnetic field $A_{\mu }$, the
equations of gravitational and electromagnetic fields may be obtained as
\begin{equation}
G_{\mu \nu }+\Lambda g_{\mu \nu }=T_{\mu \nu },  \label{GravEq}
\end{equation}
\begin{equation}
\partial _{\mu }\left( \sqrt{-g}F^{\mu \nu }F^{s-1}\right) =0,  \label{MaxEq}
\end{equation}
In the presence of nonlinear electrodynamics field, the energy-momentum
tensor of Eq. (\ref{GravEq}) is
\begin{equation}
T_{\mu \nu }=2\alpha \left[ sF_{\mu \rho }F_{\nu }^{\rho }F^{s-1}-\frac{1}{4}%
g_{\mu \nu }F^{s}\right] .  \label{TT}
\end{equation}
It is easy to show that when $s$ goes to $1$, the Eqs. (\ref{IG})--(\ref{TT}%
), reduce to the equations of magnetic brane in Einstein-standard Maxwell
gravity \cite{Dehhorizonless1}.

\section{Static magnetic branes\label{Long}}

Here we want to obtain the $d$-dimensional solutions of Eqs. (\ref{GravEq}),
(\ref{MaxEq}) which produce longitudinal magnetic fields in the Euclidean
submanifold spans by $x^{i}$\ coordinates ($i=1,...,d-3$). We will work with
the following ansatz for the metric \cite{DiasLemos02}:
\begin{equation}
ds^{2}=-\frac{\rho ^{2}}{l^{2}}dt^{2}+\frac{d\rho ^{2}}{f(\rho )}%
+l^{2}f(\rho )d\phi ^{2}+\frac{\rho ^{2}}{l^{2}}dX^{2},  \label{Met1a}
\end{equation}%
where $dX^{2}={{\sum_{i=1}^{d-3}}}(dx^{i})^{2}$ is the Euclidean metric on
the $(d-3)$-dimensional submanifold. The angular coordinates $\phi $ is
dimensionless as usual and ranges in $[0,2\pi ]$, while $x^{i}$'s range in $%
(-\infty ,\infty )$. Also, one can obtain this metric with transformations $%
t\rightarrow il\phi $\ and $\phi \rightarrow it/l$ in the horizon flat
Schwarzschild like metric, $ds^{2}=-f(\rho )dt^{2}+\frac{d\rho ^{2}}{f(\rho )%
}+\rho ^{2}d\phi ^{2}+\frac{\rho ^{2}}{l^{2}}dX^{2}$. The motivation for
this metric gauge $[g_{tt}\varpropto -\rho ^{2}$ and $(g_{\rho \rho
})^{-1}\varpropto g_{\phi \phi }]$ instead of the usual Schwarzschild gauge $%
[(g_{\rho \rho })^{-1}\varpropto g_{tt}$ and $g_{\phi \phi }\varpropto \rho
^{2}]$ comes from the fact that we are looking for a horizonless magnetic
solutions. Taking the trace of the gravitational field equation (\ref{GravEq}%
), the scalar curvature is expressed in terms of the Maxwell invariant $F$ \
and cosmological constant $\Lambda $\ as
\[
R=\frac{d\Lambda +\alpha (d-4s)F^{s}}{d-2}.
\]%
Before studying in details the field equations, we first specify the sign of
the coupling constant $\alpha $ in term of the exponent $s$ in order to
ensure a physical interpretation of our future solutions. In fact, the sign
of the coupling constant $\alpha $ in the action (\ref{IG}) can be chosen
such that the energy density, i.e. the $T_{_{\widehat{t}\widehat{t}}}$
component of the energy-momentum tensor in the orthonormal frame, is
positive
\[
T_{_{\widehat{t}\widehat{t}}}=\frac{\alpha }{2}F^{s}>0.
\]%
As a direct consequence, one can show that the Maxwell invariant $F=\frac{2}{%
l^{2}}(F_{\phi \rho })^{2}$ is positive and hence, the sign of the coupling
constant $\alpha $ should be positive, which can be set to $1$ without loss
of generality. The gauge potential is given by
\begin{equation}
A_{\mu }=h(\rho )\delta _{\mu }^{\phi }.  \label{Pot1a}
\end{equation}%
The nonlinear electromagnetic field equations (\ref{MaxEq}) reduce to
\begin{equation}
\rho (2s-1)\frac{d^{2}h(\rho )}{d\rho ^{2}}+(d-2)\frac{dh(\rho )}{d\rho }=0,
\end{equation}%
for $s\neq 1/2,(d-1)/2$, with the solution
\begin{equation}
h(\rho )=-\frac{2ql^{d-2}}{(d-3)\rho ^{(d-2s-1)/(2s-1)}}.  \label{Em2}
\end{equation}%
where $q$ is an integration constant and is related to the electromagnetic
charge. It is easy to show that the only nonzero electromagnetic field
tensor is
\begin{equation}
F_{\phi \rho }=\frac{2(2s-d+1)ql^{d-2}}{(d-3)(2s-1)\rho ^{(d-2)/(2s-1)}}.
\label{Fpr}
\end{equation}%
To find the function $f(\rho )$, one may use any components of Eq. (\ref%
{GravEq}). The simplest equation is the $\rho \rho $ component of these
equations which can be written as
\begin{equation}
(d-2)\rho f^{^{\prime }}(\rho )+(d-2)(d-3)f(\rho )+2\Lambda \rho
^{2}-2^{s}(2s-1)\rho ^{2}\left( \frac{2\left( 2s-d+1\right) ql^{d-3}}{%
(d-3)\left( 2s-1\right) \rho ^{\frac{(s-2)}{2s-1}}}\right) ^{2s}=0,
\label{rrcom}
\end{equation}%
for $s\neq 1/2,(d-1)/2$, where the prime denotes a derivative with respect
to the $\rho $ coordinate. The solution of Eq. (\ref{rrcom}), which also
satisfies all the other components of the gravitational field equations (\ref%
{GravEq}), can be written as
\begin{equation}
f(\rho )=-{\frac{2\Lambda \,\rho ^{2}}{\left( d-1\right) \left( d-2\right) }}%
+\frac{2^{s}\left( 2s-1\right) ^{2}\rho ^{2(4s-ds-1)/(2s-1)}}{\left(
d-2\right) \left( 2s-d+1\right) }\left( \,{\frac{2\left( 2s-d+1\right)
ql^{d-3}}{(d-3)\left( 2s-1\right) }}\right) ^{2s}+{\frac{2ml^{3}}{\rho ^{d-3}%
}},  \label{f(r)}
\end{equation}%
where mass parameter, $m,$ is related to integration constant.

\subsubsection{Special Cases (I): $S=(d-1)/2$ and $s=1/2$}

After some manipulations, we find that for $s=1/2$ and $(d-1)/2$, the gauge
potential $A_{\mu }$ and the charge term in the metric function differ from
previous section.

It is notable that for $s=1/2$, all components of the electromagnetic field,
$F_{\mu \nu }$, vanish and the solutions reduce to uncharged solutions.
While for $s=(d-1)/2$, there is a logarithmic function related to the charge
term of the solutions obtained as followed%
\begin{eqnarray}
h(r) &=&-\frac{2ql^{d-2}}{d-3}\ln \rho , \\
f(\rho ) &=&-{\frac{2\Lambda \,\rho ^{2}}{\left( d-1\right) \left(
d-2\right) }}+\left( \frac{2^{3/2}l^{d-3}q}{d-3}\right) ^{d-1}\frac{\ln \rho
}{\rho ^{d-3}}+{\frac{2ml^{3}}{\rho ^{d-3}}},
\end{eqnarray}

\subsubsection{Special Cases (II): $4$-dimensional solutions}

Here, with an appropriate combination of nonlinear electromagnetic field and
$4$-dimensional Einstein gravity, we constructed a class of four dimensional
magnetic string solutions which produces a longitudinal magnetic field. In $%
4 $-dimensional case the metric reduce to magnetic string coupled to
nonlinear electromagnetic field%
\begin{equation}
ds^{2}=-\frac{\rho ^{2}}{l^{2}}dt^{2}+\frac{d\rho ^{2}}{f(\rho )}%
+l^{2}f(\rho )d\phi ^{2}+\frac{\rho ^{2}}{l^{2}}dz^{2},
\end{equation}

The coordinate $z$ has the dimension of length and ranges $-\infty <z<\infty
$ and the electromagnetic field and metric function reduce to%
\begin{eqnarray}
F_{\phi \rho } &=&\left\{
\begin{array}{cc}
0 & ,s=0,1/2 \\
-\frac{2ql^{2}}{\rho } & ,s=3/2 \\
\frac{2(2s-3)ql^{2}}{(2s-1)\rho ^{2/(2s-1)}} & ,\text{Else}%
\end{array}%
\right. , \\
f(\rho ) &=&-{\frac{\Lambda \,\rho ^{2}}{3}}+{\frac{2ml^{3}}{\rho }+}\left\{
\begin{array}{cc}
0, & s=0,1/2 \\
\left( \frac{2^{3/2}l^{d-3}q}{d-3}\right) ^{d-1}\frac{\ln \rho }{\rho ^{d-3}}%
, & s=3/2 \\
\frac{2^{s}\left( 2s-1\right) ^{2}}{2\left( 2s-3\right) \rho ^{2/(2s-1)}}%
\left( \,{\frac{2\left( 2s-3\right) ql}{\left( 2s-1\right) }}\right) ^{2s},
& \text{Else}%
\end{array}%
\right. .
\end{eqnarray}

\section{\protect\bigskip Properties of the Solutions:}

\subsection{Comparing of the metric function in the Einstein-Maxwell gravity
with the Einstein nonlinear Maxwell gravity}

In this subsection, we want to investigate the effects of the nonlinearity
of the magnetic field on the metric (function). It is clear that the
nonlinearity effects on the second term of $f(\rho )$, Eq. (\ref{f(r)}),
directly. But for the study of the spacetime structure, it is better to
investigate the behavior of the metric function for small and large value of
$\rho $. The metric function of the \emph{Einstein-standard Maxwell} is
\begin{equation}
f(\rho )=-{\frac{2\Lambda \,\rho ^{2}}{\left( d-1\right) \left( d-2\right) }}%
+\frac{4ql^{d-3}\rho ^{2(3-d)}}{\left( d-2\right) (d-3)}+{\frac{2ml^{3}}{%
\rho ^{d-3}}}.  \label{fMax}
\end{equation}%
One can show that for small and large value of $\rho $ ($\rho
\longrightarrow 0$ and $\rho \longrightarrow \infty $), the dominant terms
of the metric function (\ref{fMax}) are second and first term, respectively.
It means that
\begin{eqnarray*}
\left. \lim_{\rho \longrightarrow 0}f(\rho )\right\vert _{\text{standard
Maxwell}} &\varpropto &\rho ^{2(3-d)}, \\
\left. \lim_{\rho \longrightarrow \infty }f(\rho )\right\vert _{\text{%
standard Maxwell}} &\varpropto &\rho ^{2}.
\end{eqnarray*}%
It is easy to find that these solutions (Einstein-standard Maxwell
solutions) are asymptotically AdS. But for \emph{nonlinear Maxwell field
coupled to Einstein gravity} (our case), we have
\begin{eqnarray*}
\left. \lim_{\rho \longrightarrow 0}f(\rho )\right\vert _{\text{nonlinear
Maxwell}} &\varpropto &\left\{
\begin{array}{l}
\rho ^{2(4s-ds-1)/(2s-1)}\text{ , for }\frac{1}{2}<s<(d-1)/2 \\
\rho ^{3-d}\ln \rho \text{ \ \ \ \ \ \ \ \ \ \ \ , for }s=(d-1)/2 \\
\rho ^{3-d}\text{ \hspace{2cm} , else }%
\end{array}%
\right. \\
\left. \lim_{\rho \longrightarrow \infty }f(\rho )\right\vert _{\text{
nonlinear Maxwell}} &\varpropto &\left\{
\begin{array}{l}
\rho ^{2(4s-ds-1)/(2s-1)}\text{\ , for }0<s<\frac{1}{2} \\
\rho ^{2}\text{\hspace{2.5cm} , for }s>\frac{1}{2}%
\end{array}%
\right.
\end{eqnarray*}%
It is worthwhile to mention that if $0<s<\frac{1}{2}$, the
solutions of the Einstein-nonlinear Maxwell field equations are
not asymptotically AdS and for $s>\frac{1}{2}$, the asymptotic
behavior of Einstein-nonlinear Maxwell field solutions are the
same as linear case. It is interesting that the nonlinear Maxwell
field effects on all values of $r$. There is not any restriction
on the nonlinear parameter $s$ and one can fix it for consistency
to other models.

\subsection{Singularities}

In order to study the general structure of the solutions given in Eqs. (\ref%
{Met1a}) and (\ref{f(r)}), we first look for curvature singularities. One
can find that the Kretschmann scalar, $R_{\mu \nu \lambda \kappa }R^{\mu \nu
\lambda \kappa }$, is
\[
R_{\mu \nu \lambda \kappa }R^{\mu \nu \lambda \kappa }=\left( \frac{%
d^{2}f(\rho )}{d{\rho }^{2}}\right) ^{2}+2(d-2)\left( \frac{1}{\rho }\frac{%
df(\rho )}{d{\rho }}\right) ^{2}+2(d-2)(d-3)\left( \frac{f(\rho )}{{\rho }%
^{2}}\right) ^{2}.
\]%
It is easy to show that the Kretschmann scalar $R_{\mu \nu \lambda \kappa
}R^{\mu \nu \lambda \kappa }$ diverge at $\rho =0$ and therefore one might
think that there is a curvature singularity located at $\rho =0$. However,
as we will see, the spacetime will never achieve $\rho =0$. The function $%
f(\rho )$ is negative for $\rho <r_{+}$ and positive for $\rho >r_{+}$,
where $r_{+}$ is the largest root of $f(\rho )=0$. The function $g_{\rho
\rho }$ cannot be negative (which occurs for $\rho <r_{+}$), because of the
change of signature of the metric. Thus, one cannot extend the spacetime to $%
\rho <r_{+}$. To get rid of this incorrect extension, we can introduce the
new radial coordinate $r$ as
\[
r^{2}=\rho ^{2}-r_{+}^{2}\Rightarrow d\rho ^{2}=\frac{r^{2}}{r^{2}+r_{+}^{2}}%
dr^{2}.
\]%
With this new coordinate, the metric (\ref{Met1a}) is
\begin{equation}
ds^{2}=-\frac{r^{2}+r_{+}^{2}}{l^{2}}dt^{2}+\frac{r^{2}}{%
(r^{2}+r_{+}^{2})f(r)}dr^{2}+l^{2}f(r)d\phi ^{2}+\frac{r^{2}+r_{+}^{2}}{l^{2}%
}dX^{2},  \label{Met1b}
\end{equation}%
where the coordinate $r$ and $\phi $ assume the value $0\leq r<$ $\infty $
and $0\leq \phi <2\pi $. The function $f(r)$ is now given as
\begin{equation}
f(r)=-{\frac{2\Lambda \,(r^{2}+r_{+}^{2})}{\left( d-1\right) \left(
d-2\right) }}+\frac{2^{s}\left( 2s-1\right) ^{2}\left( {\frac{2\left(
2s-d+1\right) ql^{d-3}}{(d-3)\left( 2s-1\right) }}\right) ^{2s}}{\left(
d-2\right) \left( 2s-d+1\right) (r^{2}+r_{+}^{2})^{(ds-4s+1)/(2s-1)}}+{\frac{%
2ml^{3}}{(r^{2}+r_{+}^{2})^{(d-3)/2}}},  \label{F2}
\end{equation}%
for $s\neq 0,1/2,(d-1)/2$ , when $s=0,1/2$ it is sufficient to let $q=0$ in
Eq. (\ref{F2}) and for $s=(d-1)/2$, we should substitute second term of \
Eq. (\ref{F2}) with $\frac{1}{2}\left( \frac{2^{3/2}l^{d-3}q}{d-3}\right)
^{d-1}\frac{\ln (r^{2}+r_{+}^{2})}{(r^{2}+r_{+}^{2})^{(d-3)/2}}$. The
electromagnetic field equation in the new coordinate is
\begin{equation}
F_{\phi r}=\left\{
\begin{array}{cc}
0 & ,s=0,1/2 \\
-\frac{2ql^{d-2}}{(d-3)(r^{2}+r_{+}^{2})^{1/2}} & ,s=(d-1)/2 \\
\frac{2(2s-d+1)ql^{d-2}}{(d-3)(2s-1)(r^{2}+r_{+}^{2})^{(d-2)/[2(2s-1)]}} & ,%
\text{Else}%
\end{array}%
\right. .  \label{f33}
\end{equation}%
The function $f(r)$ given in Eq. (\ref{f(r)}) is positive in the whole
spacetime and is zero at $r=0$. \ One can show that all curvature invariants
(such as Kretschmann scalar, Ricci scalar, Ricci square, Weyl square and so
on) are functions of $f^{\prime \prime }$, $f^{\prime }/r$ and $f/r^{2}$.
Since these terms do not diverge in \ the range $0\leq r<\infty $, one finds
that all curvature invariants are finite. Therefore this spacetime has no
curvature singularities and no horizons. It is worthwhile to mention that
the magnetic solutions obtained here have distinct properties relative to
the electric solutions obtained in \cite{HendiRast}. Indeed, the electric
solutions have black holes, while the magnetic do not. However, the metric (%
\ref{Met1a}) has a conic geometry and has a conical singularity at $r=0$.
For e.g., when $s\neq 0,1/2,(d-1)/2$ and using a Taylor expansion in the
vicinity of $r=0$, we can write
\begin{eqnarray}
f(r) &=&f\left. (r)\right\vert _{r=0}+\left. r\frac{df(r)}{dr}\right\vert
_{r=0}+\left. \frac{r^{2}}{2}\frac{d^{2}f(r)}{dr^{2}}\right\vert _{r=0}+....
\nonumber \\
&&  \nonumber \\
\left. f(r)\right\vert _{r=0} &=&0,\text{ \ \ \ \ \ \ \ \ \ \ \ }\left. r%
\frac{df(r)}{dr}\right\vert _{r=0}=0,  \nonumber \\
&&  \nonumber \\
\left. \frac{d^{2}f(r)}{dr^{2}}\right\vert _{r=0} &=&-\frac{2\Lambda }{d-2}%
+\left\{
\begin{array}{cc}
\left( \frac{2^{3/2}l^{d-3}q}{(d-3)r_{+}}\right) ^{d-1} & ,s=(d-1)/2 \\
\\
\frac{2^{3s}(s-1)q^{2s}l^{2s(d-3)}}{(d-3)^{2s}\left( \frac{2s-1}{2s-d+1}%
\right) ^{2s-1}r_{+}^{(ds-1)/(2s-1)}} & ,s\neq (d-1)/2%
\end{array}%
\right. .  \nonumber
\end{eqnarray}%
As a result, we can rewrite Eq. (\ref{Met1b})
\begin{equation}
ds^{2}=-\frac{r_{+}^{2}}{l^{2}}dt^{2}+\frac{2\left( \frac{d^{2}f}{dr^{2}}%
\left\vert _{r=0}\right. \right) ^{-1}}{r_{+}^{2}}dr^{2}+\frac{l^{2}}{2}%
\left( \frac{d^{2}f}{dr^{2}}\left\vert _{r=0}\right. \right) r^{2}d\phi ^{2}+%
\frac{r_{+}^{2}}{l^{2}}dX^{2},  \label{met0}
\end{equation}%
then
\[
\lim_{r\rightarrow 0}\frac{1}{r}\sqrt{\frac{g_{\phi \phi }}{g_{rr}}}%
=\lim_{r\rightarrow 0}\frac{\sqrt{(r^{2}+r_{+}^{2})}lf(r)}{r^{2}}=\frac{%
lr_{+}}{2}\left( \frac{d^{2}f(r)}{dr^{2}}\mid _{r=0}\right) \neq 1,
\]%
which clearly shows that the spacetime has a conical singularity at $r=0$
since, when the radius $r$ tends to zero, the limit of the ratio \emph{%
circumference/radius} is not $2\pi $. The conical singularity can be removed
if one identifies the coordinate $\phi $ with the period
\begin{equation}
\text{Period}_{\phi }=2\pi \left( \lim_{r\rightarrow 0}\frac{1}{r}\sqrt{%
\frac{g_{\phi \phi }}{g_{rr}}}\right) ^{-1}=2\pi (1-4\mu ),
\end{equation}%
where
\[
\mu =\frac{1}{4}\left[ 1-\frac{2}{lr_{+}}\left( \frac{d^{2}f(r)}{dr^{2}}\mid
_{r=0}\right) ^{-1}\right] .
\]%
One concludes that near the origin, $r=0$, metric (\ref{met0}) describes a
spacetime which is locally flat but has a conical singularity at $r=0$ with
a deficit angle $\delta \phi =8\pi \mu $. In four dimensional solutions, $%
\mu $ interpreted as the mass per unit length of the magnetic string.
Therefore, in higher dimensions, we may interpret $\mu $ as the mass per
unit volume of the magnetic brane.

\begin{figure}[tbp]
\epsfxsize=7cm \centerline{\epsffile{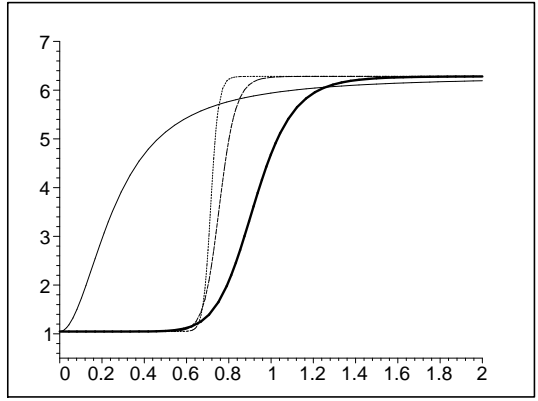} }
\caption{The deficit angle versus $q $ for $d=5$, $r_{+}=0.6$, $l=1$, and $%
s=1$ (continuous line), $s=5$ (bold line), $s=10$ (dashed line) and $s=20$
(dotted line) .}
\label{Fig1}
\end{figure}

Also, in order to investigate the effects of the nonlinearity of the
magnetic field on $\mu $, we plot the deficit angle $\delta \phi $ versus
the charge parameter $q$. This is shown in Fig. \ref{Fig1}, which shows that
the nonlinear parameter effects on the deficit angle unless for $%
q\rightarrow 0$ and $q\rightarrow \infty $.

Of course, one may ask for the completeness of the spacetime with $r\geq 0$
(or $\rho \geq r_{+}$). It is easy to see that the spacetime described by
Eq. (\ref{Met1b}) is both null and timelike geodesically complete as in the
case of four-dimensional standard Maxwell solutions \cite{DiasLemos02,Hor}.
In fact, one can show that every null or timelike geodesic starting from an
arbitrary point can either extend to infinite values of the affine parameter
along the geodesic or end on a singularity \cite{DB07}.

\section{Spinning Magnetic Branes\label{Rot}}

Now, we want to endow our spacetime solutions (\ref{Met1a}) with a global
rotation. At first, we consider the solutions with one rotation parameter.
In order to add angular momentum to the spacetime, we perform the following
rotation boost in the $t$-$\phi $ plane
\begin{equation}
t\mapsto \Xi t-a\phi \ \ \ \ \ \ \ \ \ \ \phi \mapsto \Xi \phi -\frac{a}{%
l^{2}}t,  \label{tphi}
\end{equation}%
where $a$ is the rotation parameter and $\Xi =\sqrt{1+a^{2}/l^{2}}$.
Substituting Eq. (\ref{tphi}) into Eq. (\ref{Met1a}) we obtain
\begin{equation}
ds^{2}=-\frac{r^{2}}{l^{2}}\left( \Xi dt-ad\phi \right) ^{2}+\frac{dr^{2}}{%
f(r)}+l^{2}f(r)\left( \frac{a}{l^{2}}dt-\Xi d\phi \right) ^{2}+\frac{r^{2}}{%
l^{2}}dX^{2},  \label{Metrot}
\end{equation}%
where $f(r)$ is the same as $f(r)$ given in Eq. (\ref{F2}). The non
vanishing electromagnetic field components become
\begin{equation}
F_{rt}=-\frac{a}{\Xi l^{2}}F_{r\phi }=\frac{2ql^{d-4}a}{(d-3)}\times \left\{
\begin{array}{cc}
0 & ,s=0,1/2 \\
-(r^{2}+r_{+}^{2})^{-1/2} & ,s=(d-1)/2 \\
\frac{(2s-d+1)}{(2s-1)(r^{2}+r_{+}^{2})^{(d-2)/[2(2s-1)]}} & ,\text{Else}%
\end{array}%
\right.
\end{equation}%
Because of the periodic nature of $\phi $, the transformation (\ref{tphi})
is not a proper coordinate transformation on the entire manifold. Therefore,
the metrics (\ref{Met1b}) and (\ref{Metrot}) can be locally mapped into each
other but not globally, and so they are distinct \cite{Stachel}. Again, this
spacetime has no horizon and curvature singularity, However, it has a
conical singularity at $r=0$.

Second, we study the rotating solutions with more rotation parameters. The
rotation group in $d$ dimensions is $SO(d-1)$ and therefore the number of
independent rotation parameters is $[(d-1)/2]$, where $[x]$ is the integer
part of $x$. We now generalize the above solution given in Eq. (\ref{Met1b})
with $k\leq \lbrack (d-1)/2]$ rotation parameters. This generalized solution
can be written as
\begin{eqnarray}
ds^{2} &=&-\frac{r^{2}}{l^{2}}\left( \Xi dt-{{\sum_{i=1}^{k}}}a_{i}d\phi
^{i}\right) ^{2}+f(r)\left( \sqrt{\Xi ^{2}-1}dt-\frac{\Xi }{\sqrt{\Xi ^{2}-1}%
}{{\sum_{i=1}^{k}}}a_{i}d\phi ^{i}\right) ^{2}  \nonumber \\
&&+\frac{dr^{2}}{f(r)}+\frac{r^{2}}{l^{2}(\Xi ^{2}-1)}{\sum_{i<j}^{k}}%
(a_{i}d\phi _{j}-a_{j}d\phi _{i})^{2}+\frac{r^{2}}{l^{2}}dX^{2},
\label{Metr5}
\end{eqnarray}%
where $\Xi =\sqrt{1+\sum_{i}^{k}a_{i}^{2}/l^{2}}$, $dX^{2}$ is the Euclidean
metric on the $(d-k-2)$-dimensional submanifold with volume $V_{d-k-2}$ and $%
f(r)$ is the same as $f(r)$ given in Eq. (\ref{f(r)}). The non-vanishing
components of electromagnetic field tensor are
\begin{equation}
F_{rt}=-\frac{(\Xi ^{2}-1)}{\Xi a_{i}}F_{r\phi ^{i}}=\frac{2ql^{d-3}\sqrt{%
\Xi ^{2}-1}}{(d-3)}\times \left\{
\begin{array}{cc}
0 & ,s=0,1/2 \\
-(r^{2}+r_{+}^{2})^{-1/2} & ,s=(d-1)/2 \\
\frac{(2s-d+1)}{(2s-1)(r^{2}+r_{+}^{2})^{(d-2)/[2(2s-1)]}} & ,\text{Else}%
\end{array}%
\right. .
\end{equation}%
One can see that rotation does not change the metric function $f(r)$, and
therefore the metric in the transformed coordinate systemIn the remaining
part of this section, we compute the conserved quantities of the solutions.
In general, the conserved quantities are divergent when evaluated on the
solutions. A systematic method of dealing with this divergence for
asymptotically AdS solutions of Einstein gravity is through the use of the
counterterms method inspired by the anti-de Sitter conformal field theory
(AdS/CFT) correspondence \cite{Mal}. One can use the Brown-York definition
of the stress-energy tensor \cite{BY} to construct a divergence-free
stress-energy tensor
\begin{equation}
T^{ab}=\frac{1}{8\pi }\left[ (K^{ab}-K\gamma ^{ab})+\frac{d-2}{l}\gamma ^{ab}%
\right] ,  \label{Stres}
\end{equation}%
where $K^{ab}$ is the extrinsic curvature of the boundary, $K$ is its trace
and $\gamma ^{ab}$ is the induced metric of the boundary.

To compute the conserved charges of the spacetime, we choose a spacelike
surface $\mathcal{B}$ in $\partial \mathcal{M}$ with metric $\sigma _{ij}$,
and write the boundary metric in Arnowitt-Deser-Misner form:
\begin{equation}
\gamma _{ab}dx^{a}dx^{a}=-N^{2}dt^{2}+\sigma _{ij}\left( d\varphi
^{i}+V^{i}dt\right) \left( d\varphi ^{j}+V^{j}dt\right) ,
\end{equation}%
where the coordinates $\varphi ^{i}$ are the angular variables
parameterizing the hypersurface of constant $r$ around the origin, and $N$
and $V^{i}$ are the lapse and shift functions respectively. When there is a
Killing vector field $\mathcal{\xi }$ on the boundary, then the quasilocal
conserved quantities associated with the stress tensors of Eq. (\ref{Stres})
can be written as
\begin{equation}
\mathcal{Q}(\mathcal{\xi )}=\int_{\mathcal{B}}d^{d-2}\varphi \sqrt{\sigma }%
T_{ab}n^{a}\mathcal{\xi }^{b},  \label{charge}
\end{equation}%
where $\sigma $ is the determinant of the metric $\sigma _{ij}$, and $n^{a}$
is the timelike unit normal vector to the boundary $\mathcal{B}$\textbf{.}
In the context of counterterm method, the limit in which the boundary $%
\mathcal{B}$ becomes infinite ($\mathcal{B}_{\infty }$) is taken, and the
counterterm prescription ensures that the action and conserved charges are
finite. For our case, horizonless rotating spacetimes, the first Killing
vector is $\xi =\partial /\partial t$ and therefore its associated conserved
charge of the brane is the mass per unit volume $V_{d-k-2}$ calculated as
\begin{equation}
M=\int_{\mathcal{B}}d^{d-2}x\sqrt{\sigma }T_{ab}n^{a}\xi ^{b}=\frac{(2\pi
)^{k}}{4}\left[ (d-1)(\Xi ^{2}-1)+1\right] m.  \label{Mas}
\end{equation}%
The second class of conserved quantities are the angular momentum per unit
volume $V_{d-k-2}$ associated with the rotational Killing vectors $\zeta
_{i}=\partial /\partial \phi ^{i}$ which may be calculated as
\begin{equation}
J_{i}=\int_{\mathcal{B}}d^{d-2}x\sqrt{\sigma }T_{ab}n^{a}\zeta _{i}^{b}=%
\frac{(2\pi )^{k}}{4}(d-1)m\Xi a_{i}.  \label{Ang}
\end{equation}
One can find that the mass and angular momentum, Eqs. (\ref{Mas}) and (\ref%
{Ang}), do not depend on the nonlinear parameter $s$. It is a matter of
calculation to show that the mass and angular momentum per unit volume of
the solutions for $0 < s < \frac{1}{2}$ is the same as asymptotic AdS case.

Next, we calculate the electric charge of the solutions. To determine the
electric field we should consider the projections of the electromagnetic
field tensor on special hypersurfaces.The electric charge per unit volume $%
V_{d-k-2}$ can be found by calculating the flux of the electromagnetic field
at infinity, yielding
\begin{equation}
Q=\frac{(2\pi )^{k}\sqrt{\Xi ^{2}-1}}{2}\times \left\{
\begin{array}{cc}
\frac{s}{8^{s}(2s-d+1)}\left( \frac{8(2s-d+1)(2s-1)ql}{d-3}\right) ^{2s-1} &
,s\neq (d-1)/2 \\
\\
2^{(3d-11)/2}\left( \frac{ql}{d-3}\right) ^{d-2} & ,s=(d-1)/2%
\end{array}%
\right. .  \label{elecch}
\end{equation}%
Note that the electric charge is proportional to the rotation parameter, and
is zero for the case of static solutions.

\section{Conformally Invariant Maxwell Source}

In many papers, straightforward generalization of the Maxwell field to
higher dimensions one essential property of the electromagnetic field is
lost, namely, conformal invariance. As one can see the clue of the conformal
invariance lies in the fact that we have considered power of the Maxwell
invariant, $F=F_{\mu \nu }F^{\mu \nu }$. Here we want to justify $s$, such
that the electromagnetic field equation be invariant under conformal
transformation ($g_{\mu \nu }\longrightarrow \Omega ^{2}g_{\mu \nu }$ and $%
A_{\mu }\longrightarrow A_{\mu }$). The idea is to take advantage of the
conformal symmetry to construct the analogues of the four dimensional
Reissner-Nordstr\"{o}m solutions in higher dimensions.

Consider the Lagrangian of the form $L(F)$, where $F=F_{\mu \nu}F^{\mu \nu}$%
. It is easy to show that for this form of Lagrangian in $d$-dimensions, $%
T_{\mu}^{\mu}\propto \left[F\frac{dL}{dF}-\frac{d}{4 }L\right]$; so $%
T_{\mu}^{\mu}=0$ implies $L(F)=Constant\times F^{\frac{d}{4} } $. For our
case $L(F)\propto F^{s}$, and subsequently, $s=\frac{d}{4}$. It is
worthwhile to mention that Since $d\geq 4$ and therefore $s\geq 1$, one can
show that the conformally invariant Maxwell source solutions are
asymptotically AdS in arbitrary dimensions $d\geq4$.

\section{Closing Remarks}

In this paper, we investigated the effects of the nonlinearity of the
electromagnetic fields on the properties of the spacetime by finding a new
class of magnetic solutions in Einstein gravity in the presence of nonlinear
electromagnetic field. This class of solutions yields an $d$-dimensional
spacetime with a longitudinal nonlinear magnetic field [the only nonzero
component of the vector potential is $A_{\phi }(r)$] generated by a static
magnetic brane. Also, we compared the asymptotic behavior of metric
functions for linear and nonlinear cases and showed that the nonlinear case
is not asymptotic AdS for special range of nonlinear parameter, ($0<s<\frac{1%
}{2}$). We found that these solutions have no curvature singularity and no
horizons, but have conic singularity at $r=0$ with a deficit angle $\delta
\phi$. We found that this deficit angle is different from the deficit angle
of spacetimes of Einstein gravity in the presence of Maxwell field for large
value of charge parameter. Next, we endow the spacetime with rotation. For
the spinning brane, when the rotation parameters are nonzero, the brane has
a net electric charge density which is proportional to the magnitude of the
rotation parameters given by $\sqrt{\Xi ^{2}-1}$. We also applied the
counterterm method in order to calculate the conserved quantities of the
spacetime, and found that the mass and angular momentum per unit volume of
the spacetime do not depend on the nonlinear parameter. Finally, we
considered the conformally invariant Maxwell source of the solutions and
found that the nonlinear parameter restricted to $s=\frac{d}{4}$, and
spacetime is asymptotic AdS for arbitrary dimensions $d\geq4$.

\begin{acknowledgements}
This work has been supported financially by National Elite
Foundation of Iran.
\end{acknowledgements}

\end{document}